\documentclass[aps, preprintnumbers,notitlepage, twocolumn]{revtex4-1}
\usepackage{amsmath, amsthm, amssymb}
\usepackage{mathtext}
\usepackage{pifont}
\usepackage{dsfont}
\usepackage{amscd}
\usepackage{amsfonts}
\usepackage{textcomp}
\usepackage{enumerate}
\usepackage[bookmarks=false,colorlinks=true,linkcolor={red},pdfstartview={XYZ null null 1.22}]{hyperref}
\usepackage{float}
\usepackage{mathpazo}

\fontfamily{palatino-ttf}

\usepackage[pdftex]{graphicx}
\usepackage[usenames,dvipsnames]{color}

\usepackage{mathpazo}
\newcommand{\be}{\begin{equation}}
\newcommand{\ee}{\end{equation}}
\newcommand{\ben}{\begin{enumerate}}
\newcommand{\een}{\end{enumerate}}

\newcommand{\half}{\frac{1}{2}}
\newcommand{\chn}{{\cal N}}

\newcommand{\pdit}{{\gamma^{(d)}}}
\newcommand{\pditp}{{\gamma^{(d)}_+}}
\newcommand{\pditpm}{{\gamma^{(d)}_\pm}}

\newtheorem{thm}{Theorem}[section]
\newtheorem{lemma}{Lemma}

\newtheorem{examp}[thm]{Example}

\newtheorem{obs}[thm]{Observation}

\fontfamily{palatino-ttf}

\usepackage[margin=2cm]{geometry}
\def\tr{{\rm Tr}}
\def\ID{\mathbb{1}}

\def\ra{\rangle}
\def\la{\langle}
\def\>{\rangle}
\def\<{\langle}
\def\a{\alpha}

\def\squareforqed{\hbox{\rlap{$\sqcap$}$\sqcup$}}
\def\qed{\ifmmode\squareforqed\else{\unskip\nobreak\hfil
\penalty50\hskip1em\null\nobreak\hfil\squareforqed
\parfillskip=0pt\finalhyphendemerits=0\endgraf}\fi}

\providecommand{\norm}[1]{\lVert#1\rVert}

\begin{document}

\title{Hybrid Zero-capacity Channels}

\author{Sergii Strelchuk}
\affiliation{Department of Applied Mathematics and Theoretical Physics, University of Cambridge, Cambridge CB3 0WA, U.K.}
\author{Jonathan Oppenheim}
\affiliation{Department of Applied Mathematics and Theoretical Physics, University of Cambridge, Cambridge CB3 0WA, U.K. \\ University College of London, Department of Physics \& Astronomy, London, WC1E 6BT and London Interdisciplinary Network for Quantum Science}

\begin{abstract}
There are only two known kinds of zero-capacity channels. The first kind produces entangled states that have positive partial transpose, and the second one -- states that are cloneable.  We consider the family of 'hybrid' quantum channels, which lies in the intersection of the above classes of channels and investigate its properties. It gives rise to the first explicit examples of the channels, which create bound entangled states that have the property of being cloneable to the arbitrary finite number of parties. Hybrid channels provide the first example of highly cloneable binding entanglement channels, for which  known superactivation protocols must fail -- superactivation is the effect where two channels each with zero quantum capacity having positive capacity when used together. We give two methods to construct a hybrid channel from any binding entanglement channel. We also find the low-dimensional counterparts of hybrid states -- bipartite qubit states which are extendible and possess two-way key.
\end{abstract}
\maketitle
\section{introduction}
In classical information theory, channels which cannot convey information are boring. The only such channel which has this property is the
one where there is no correlation between the input and the output.

In contrast, in quantum information theory, such channels (called zero-capacity channels), have a very rich and ill-understood structure.
There is a wide and largely unexplored variety of channels which cannot reliably send quantum information.  We don't yet know how to characterize such
channels, and thus far, there are only two methods known to determine if a channel has zero capacity.  One criterion, is if the channel produces
states which have positive partial transpose (PPT)~\cite{peres_separabilitycr, horodecki_separability_1996}, in which case the channel produces states
which cannot be distilled into pure state entanglement~\cite{horodecki_binding_2000}.  A second criterion, is if the channel produces
states which would lead to cloning~\cite{devetak_capacity_2005} -- i.e. imagine that the states $\psi^{ABE}$  produced by the channel with the share $A$ held by the
sender, $B$, the share produced by the output of the channel, and $E$, the environment.  If the channel has the structure that any state at $B$ could
be recreated at $E$, then we know that the channel must have zero-capacity because if arbitrary states could be sent to $B$, then they would be cloned
at $E$, which we know to be impossible~\cite{wootters_single_1982}.
 
We believe that there are other kinds of zero-capacity channels, for example, the channels which produce an equal mixture of
the Bell states, tensored with separable hiding states as in~\cite{horodecki_general_2009}, although there is no proof of this.  
However, even if we restrict our attention to the two classes of known
zero-capacity channels, our current understanding is woefully inadequate.  For example, recently, it was shown that the lack of capacity of these
two classes of channels is only the beginning of the story.  Although each class of channel individually has zero capacity, if they 
are used jointly, they are able to convey quantum information to the receiver.  It is as if $0+0=1$.  
This effect, termed {\it superactivation}, was discovered by
 Smith and Yard~\cite{smith_superact} for two zero-capacity quantum channels: one, being the PPT quantum channel
 which can produce private keys~\cite{horodecki_general_2009}.  The second channel is from the class of cloning channels --
symmetric channels which create states $\psi^{ABE}$ which are unchanged after switching $B$ and $E$.  We say that the resulting
reduced state $\rho^{AB}$ is 2-extendible, because they have the property of one of the subsystems being cloneable, i.e. we can make a copy of $B$ on a second system $E$ (in this case, just the state $\psi^{ABE}$). If we can make $k-1$ copies of the 
state on $B$, then we say that the state is $k$-extendible.

The kinds of cloning channels which can be used in superactivation have been extended in~\cite{brandao_when_2012}, but
it is still an important open question as to what sorts of channels and what combinations of them can be superactivated.  At the moment however,
we do not even know whether all channels that produce PPT bound entangled states can be superactivated. We make progress in showing the opposite, 
by focusing our attention on the special set of channels that produce states, which are both PPT bound entangled and simultaneously cloneable. 
Since superactivation using these two classes of channels requires one channel from each class, channels which are both PPT and cloneable,
would only be superactivatable if a third class of zero-capacity channels existed.

The set of states which are both 2-extendible and PPT was shown to be non-empty~\cite{doherty_complete_2004}, but the example together with the proof that this state has the above properties was complicated and unintuitive. In our paper, we construct some simple examples, and investigate the overlap between the two classes of channels. Moreover, we investigate the overlap of a much broader set of channels -- that are not merely $2$-extendible, but $k$-extendible for any $k\ge2$, and produce the first explicit constructive example of the channels from this set for each $k$. We term them {\it hybrid} channels. Previously, for $k\ge3$ such channels were only proven to exist~\cite{doherty_complete_2004}, but no examples were known.  Here, we exhibit a method which 
can produce an entangling, $k-extendible$ PPT channel, starting from any entangling PPT channel.

It will be easier to talk about quantum channels in terms of states associated with them by the virtue of Choi-Jamioklowski (CJ) isomorphism, which imparts the properties of the states produced by the channel to those of its CJ state.

The problems of characterization of the set of bound entangled states as well as the set of $k$-extendible states is currently open. The latter problem is closely related to the problem of separability of quantum states, as the set of bound entangled states with PPT and that of extendible states approach the set of separable states under certain asymptotic conditions. Such lack of structure in describing both of the sets makes constructing hybrid channels interesting and challenging in the same time.

Arguably, one of the strongest tests for separability is the one introduced by Doherty et al.~\cite{doherty_complete_2004}. There, authors check for the presence of the certain necessary conditions held by all separable states -- the existence of the $k$-symmetric extensions with PPT -- by solving a semidefinite program. The test itself consists of a sequence of steps performed in succession: at $i$-th step we attempt to construct the $i$-extension of the quantum state by solving a semidefinite program. If at some step $i_0$ we fail to construct the $i_0$-extension for the original state, then we conclude that the state is entangled and provide an entanglement witness for it. This test is known to be complete in the sense that running it for $n$ rounds as $n\to\infty$ we are guaranteed that if the state is entangled we will stop after a finite number of rounds. However, the size of the semidefinite program grows exponentially with the number of extensions we want to construct.

Symmetric extendibility appears in numerous other important applications. It proved to be a useful tool for analysis of the protocols that distill secure key from quantum correlations~\cite{myhr_symmetric_2009}. Procedures which increase security in such protocols inherently depend on the ability to break the symmetric extension, as failure to do so could result in adversary holding one of the extensions thereby compromising the security of the protocol. At the moment, there exist no criterion for the bipartite state to be extendible, however some partial results that provide sufficient criteria have been discovered~\cite{myhr_spectrum_2009}.

Motivated by the CJ states of different zero-capacity channels used in the original superactivation scenario, in Section~\ref{beconstruct} we introduce a new one-parameter family of so-called $\it{hybrid}$ states, which simultaneously possesses the properties of the two channels in the superactivation example. The CJ state of the first channel has PPT and is bound entangled, and that of the other is $2$-extendible, which can be made into $k$-extendible by a simple manipulation. Hybrid states and hybrid channels get their name for incorporating those crucial properties of both of the channels.
This family has an entirely different structure comparing to the one introduced in~\cite{doherty_complete_2004}, where authors provide the numerical example of a $2$-extendible entangled state with PPT and give the proof of existence of the entangled states with PPT which pass the $k$-th test. We provide a simple and explicit scalable construction of $k$-extendible bound entangled states with PPT, which for any fixed $k$ pass all steps of the separability test up to $(k+1)$st. The proof that our family of states is entangled as well as the distillation protocol are interesting in its own right, as we equip parties with such non-entangling resources as backwards unidirectional classical communication together with a backwards $50\%$ erasure channel. Following the construction, we investigate the properties of these states, and segregate a family of hybrid states which become more like separable states in the sense that their degree of extendibility increases with the dimension, yet they are far from the set of all separable states. Finally, we show how to make our family of hybrid states, which are extendible with respect to one of the subsystems into hybrid states, which are extendible with respect to both of the subsystems.

In Section~\ref{apps} we show that hybrid channels constitute the examples of channels, which produce bound entangled states, that cannot be activated by any of the known protocols known to date.
Finally, in Section~\ref{classanalogue}, we discuss low dimensional hybrid states that are suitable for generating in the lab. We explore which of the features of bipartite hybrid states remain when we constrain the dimension of the Hilbert space to be $2\otimes2$. It turns out that despite the low dimension, there exist quantum states, which are reminiscent of the hybrid states, having two seemly antithetical properties of being 2-extendible and having two-way distillable secret key. These qubit states present examples of $2\otimes2$ systems which are the analogues of the hybrid states in the sense that they demonstrate relationship between extendibility and classical key in the absence of PPT which is similar to that of extendibility and bound entanglement. These states, being low dimensional and therefore more feasible to create in the lab may be of the interest experimenters seeking to incorporate them in a variety of key distribution protocols.
 
\section{Bound entangled k-extendible states}\label{beconstruct}
In this section we construct a family of states, which have PPT and are $k$-extendible by a particular composition of the states, each with one of the two properties. Depending on which state we take as a starting point for our construction we get two different families with both of the properties. 

In the first case we start with the CJ state of the erasure channel. Then, we modify it by replacing the singlet by the CJ state of the binding entanglement channel, which is bound entangled state with PPT, to obtain the state, which retains the property of being $k$-extendible and in addition becomes PPT. 
In the second case we start with the CJ state of a particular binding entanglement channel, and further replace part of the former with the CJ states of the erasure channel. Again, we obtain a bound entangled state with PPT.

We will further consider two quantum channels. 
The first one, denoted as ${\cal N}_{\phi}$, is a channel which produces bound entangled states $\phi^{AB}$ with maximally mixed reductions. Its CJ state is PPT. The second channel is the erasure channel $\chn_e^{p}(\rho) = p\rho+(1-p)|e\ra\la e|,$
which with probability $p$ faithfully transmits the input state to the receiver and with probability $1-p$ outputs a flag signalling that erasure took place.

In the first and second subsections, we show two ways to construct the state with the properties above. Then, in the following subsection, we study their properties of these families, and in the last part we provide the way for a more flexible construction that admits extensions on both of the parties.

\subsection{Mixing erasure-like states}
We turn to constructing the $k$-extendible states that have the form similar to that of the erasure channel. Consider the CJ state of the $\chn_e^{\frac{1}{k}}$:
\be
\theta_+^{AB} = \frac{1}{k}\Phi_+^{AB} + \frac{k-1}{k}\mathbb{1}^{A}\otimes\sigma^{B},
\ee where $\sigma^B=|e\ra\la e|$ is the erasure flag. 
Now, we construct the higher-dimensional analogue of $\theta_+^{AB}$, inserting instead of the maximally entangled state any bipartite bound entangled state with PPT $\phi^{AB}$ and $\tr_A\phi^{AB}=\ID^B$, $\tr_B\phi^{AB}=\ID^A$, where the subsystems $AB$ are trivially extended to the larger Hilbert space $AA'BB'$:
\be\label{erlike}
\rho^{ABA'B'} = \frac{1}{k}\phi^{AA'BB'}+\frac{k-1}{k}\mathbb{1}^{AA'}\otimes\sigma^{BB'}.
\ee
The quantum channel, whose CJ state has the form of $\rho^{ABA'B'}$, is an erasure-like channel, in the sense that applying it with probability $\frac{1}{k}$ results in Alice and Bob sharing a bound entangled state $\phi^{AA'BB'}$, which has PPT, and the other half of the time Bob receives the erasure flag $\sigma^{BB'}$ (encoded in subspace $B^{'}$), orthogonal to the support of the $\phi^{AA'BB'}$. The following Lemma shows that the state~\eqref{erlike} has indeed the desired properties, being the first constructive example of a bound entangled state with PPT, which is extendible to arbitrary many parties:
\begin{lemma}\label{distillation}
The state $\rho^{ABA'B'}$ defined in~\eqref{erlike} is $k$-extendible with respect to the subsystem $BB'$, has PPT and is bound entangled.
\end{lemma}
{\textbf{Proof:}} 

To show that the state is $k$-extendible, we first construct 2-extension of $\rho^{AA'BB'}$ with respect to the $BB'$ subsystem:
\be
\rho^{AA'BB'EE'} = \half\rho^{AA'BB'}\otimes\sigma^{EE'} + \half\rho^{AA'EE'}\otimes\sigma^{BB'}.
\ee
Noting that $\tr_{BB'} \rho^{AA'BB'}=\ID^{AA'}$, we have $\rho^{AA'BB'}\cong\tr_{BB'} \rho^{AA'BB'EE'}\cong\tr_{EE'}  \rho^{AA'BB'EE'}$, where the corresponding reductions denote different subsystems isomorphic to each other. Therefore, the state is 2-extendible. In the same way we construct the $k$-extension of~\eqref{erlike}:
\begin{align}
\rho^{AA'\overline{BB'}} = \frac{1}{k}\sum_{i=1}^{k}\rho^{AA'B_iB'_i}\otimes\sigma^{\overline{BB'}\backslash B_iB'_i},
\end{align}
where $\overline{BB'}=B_1B_1^{'}...B_kB_k^{'}$, and $\overline{BB'}\backslash B_iB'_i$ denotes the exclusion of $B_iB'_i$ from $\overline{BB'}$.
\\

To show that $\rho^{AA'BB'}$ has PPT it suffices to show that each of the summands has PPT. The state $\phi^{AA'BB'}$ has PPT by definition, and the second summand of $\rho^{AA'BB'}$ represents a separable state, hence it has PPT.\\
Finally, we need to show that the state $\rho^{ABA'B'}$ is bound entangled. Consider two parties Alice and Bob each holding subsystems $AA'$ and $BB'$ and communicating over a classical channel from Bob to Alice. Bob performs a measurement $M=\{M_0, M_1\}$ with $M_0=\ID-|e\ra\la e|, M_1=|e\ra\la e|$ on $B'$, where $|e\ra$ is the erasure flag and tells Alice the outcome of the measurement over the classical channel from Bob to Alice. When $M_1$ is measured, then they abort the protocol. Otherwise, they know that they share $\phi^{AA'BB'}$, which is bound entangled. Given that parties cannot create entanglement using backwards classical communication channel alone this shows that the original state is entangled. $\qed$

In order for the state~\eqref{erlike} to be useful in the protocols, which require entanglement distillation from the shared quantum state, e.g. proving to third party that they indeed share the state which has some bound entanglement, Alice and Bob need an additional resource, because classical communication channel from Alice to Bob is insufficient to distill entanglement from the bound entangled state. One might think that equipping the parties with some auxiliary communication resource may be of some help, but clearly this must be a resource, using which the parties are incapable to create entanglement, but only to distil it from their shared state. As we show below, there exist a special class of bound entangled states with PPT and auxiliary communication resource in the form of the $50\%$ erasure channel from Bob to Alice, which enables the parties to distill pure state entanglement from their shared state. 

To overcome the limitation, we consider quantum states, which contain $d$ bits of secrecy. They are called private dits, pdits, or twisted ebits and have the generic form~\cite{horodecki_general_2009}:
\be\label{pdit}
\gamma^{(d)}=UP^{AB}_+\otimes \sigma^{A^{'}B^{'}}U^\dagger,
\ee
where $U= \sum_{i,j=0}^{d-1} |ij\rangle\langle ij|^{AB} \otimes U_{ij}$ is a controlled unitary operation termed {\textit{twisting}} (with arbitrary unitaries 
$U_{ij}$); $P_{+}^{AB}$ is the projector onto a $d$ dimensional maximally entangled state $\Phi_+^{AB}=\frac{1}{\sqrt{d}}\sum_{i=0}^{d-1}|ii\rangle^{AB}$; and $\sigma^{A^{'}B^{'}}$ is an arbitrary state called the ``shield'' subsystem of dimension $d'$, for its presence protects private correlations. In the case when $d=2$ we will call it a pbit. Parties that 
have $A$ and $B$ subsystems of a pdit (known as the ``key'') 
can extract $\log_2d$ ebits by performing $U^\dagger$ if one of them possesses the shield $A'B'$ in its entirety. However, when
the shield is split between the two parties, it can be impossible to perform the untwisting using only local operations, and there exist
states which are arbitrarily close to pdits, yet no ebits can be produced from them, as they have PPT. 
For the purposes of our task we construct pbit to be of a particular form~\cite{horodecki_general_2009}:
\begin{align}\label{pptpbit}
\pditpm^{AA'BB'}&=\frac{p}{2}(\Phi_\pm^{AB}\otimes\tau_1^{A'B'}+ \Phi_\mp^{AB}\otimes\tau_2^{A'B'})\\
&+ (1-p)\omega_{sep}^{AB}\otimes\tau_2^{A'B'},\nonumber
\end{align}
Here $p\in(\frac{1}{4},\frac{1}{3})$, and $\tau_1^{A'B'}=\frac{1}{2^k}\left(\rho_s+\rho_a\right)^{\otimes k}, \tau_2^{A'B'}=\rho_s^{\otimes k}$ denote Werner hiding states with $\rho_s=\frac{2}{d^2+d}P_{s}, \rho_a=\frac{2}{d^2-d}P_{a}$ with $P_{sym},P_{as}$ denoting projectors on symmetric and asymmetric subspaces respectively~\cite{Werner1989}, and 
\be\label{sepnoise}
\omega_{sep}^{AB}=\half (|01\rangle\langle 01|+|10\rangle\langle10|)^{AB}.
\ee
Our construction works equally well with $\pditpm$, and for definiteness, we pick $\pditp$. This results in the state:
\be\label{pditkext}
\rho^{AA'BB'}_+ = \frac{1}{k}\pditp^{AA'BB'}+\frac{k-1}{k}\mathbb{1}^{AA'}\otimes\sigma^{BB'}.
\ee

It is easily seen that $\rho^{AAB'B'}_+$ has all the properties of state~\eqref{erlike} according to Lemma~\ref{distillation}. Having introduced the state, we also allow backwards quantum communication between the parties in the form of the $50\%$ erasure channel {\it from} Bob to Alice denoted as ${\chn_{e,B\to A}^{0.5}}$. These auxiliary resources alone are not sufficient to distill entanglement, as the quantum and classical capacity of the erasure channel are zero, and one cannot create entanglement using local operations and classical communication. This will not be the case, however, if we allow for the forward classical communication from Alice to Bob. In order to distill entanglement from the quantum state, Alice and Bob perform the following three-step protocol on the state $\rho^{AA'BB'}_+ $: 
\begin{enumerate}
\item Bob measures $M$ on the $B'$ subsystem and sends the classical outcome to Alice. If he gets outcome $M_1$, then they abort the protocol. Otherwise, they proceed to step 2.
\item Bob uses ${\chn_{e,B\to A}^{0.5}}$ to send the $B'$ subsystem to Alice. With probability $\frac{1}{k}$ he succeeds.
\item Alice performs the untwisting operation, correctly identifying $\Phi_\pm^{AB}$. She then performs one of the correction operations $\{\sigma_z, \mathbb{1}\}$ to ensure that the final state is $\Phi_+^{AB}$.
\end{enumerate}

At the end of the protocol, with probability $p_s=\frac{1}{8k}$, Alice and Bob share a maximally entangled state $\Phi_+^{AB}$. 

\subsection{Mixing pbit-like states}

This time we construct the hybrid states by starting from the construction of the PPT pbit~\eqref{pptpbit}. Then we substitute the CJ states of the erasure channels $\chn_e^{\frac{1}{k}}$, $k\ge 2$:
\begin{align}
e^{AB}_+ &= \frac{1}{k} \Phi_+^{AB} + \frac{k-1}{k}\mathbb{1}^A\otimes\sigma^B,\label{eplus}\\
e^{AB}_- &= \frac{1}{k} \Phi_-^{AB} + \frac{k-1}{k}\mathbb{1}^A\otimes\sigma^B
\end{align}
 instead of the subsystems that contain the singlet to ensure $k$-extendibility. The resulting state has the form:

\begin{align}\label{next}
\eta^{AA'BB'}_\alpha=&\frac{p}{2}([\alpha\Phi_+^{AB}+ (1-\alpha)\mathbb{1}^A\otimes\sigma^B] \otimes\tau_1^{A'B'}\\ \nonumber&+ [\alpha\Phi_-^{AB}+ (1-\alpha)\mathbb{1}^A\otimes\sigma^B]\otimes\tau_2^{A'B'})\\ &+ \alpha(1-p)\omega_{sep}^{AB}\otimes\tau_2^{A'B'},\nonumber
\end{align}
where $\alpha=\frac{1}{k}$, and $\omega_{sep}^{AB}$ is the same as in~\eqref{sepnoise}. To our knowledge, the state~\eqref{next} together with~\eqref{erlike} represent the first explicit examples of bound entangled states which pass the hierarchy of the separability criteria introduced by Doherty et al.\cite{doherty_complete_2004} up to any given level. 

We prove that this state has the same properties as~\eqref{erlike}:
\begin{lemma}\label{jonostatelemma}
The state $\eta^{AA'BB'}_\alpha$ is $k$-extendible with respect to Bob's subsystem ($BB'$), has PPT and is bound entangled.
\end{lemma}
{\textbf{Proof:}} The state~\eqref{next} is $k$-extendible as each $e^{AB}_\pm$ is $k$-extendible by construction and $\omega_{sep}^{AB}$ is separable, hence extendible to an arbitrary number of parties.

To see that the state has PPT we notice that by simple regrouping of the terms we can write it as:
\begin{align}
\eta^{AA'BB'} = \alpha\gamma^{AA'BB'}_+ + \left(1-\alpha\right)\zeta_{sep},
\end{align}
where $\gamma^{AA'BB'}_+$ is the same as in~\eqref{pptpbit}, which has PPT, and the remaining summands are separable, hence arbitrarily extendible.

To show that the state is bound entangled, it is sufficient to repeat the proof of Lemma~\eqref{distillation}, with the only difference that Bob performs a measurement $M=\{M_0, M_1\}$ on $B$.
 $\qed$
 
We will henceforward refer to both families of introduced $k$-extendible states as $k$-hybrid states.
\subsection{Properties of k-hybrid states}
One question of interest is to investigate the bounds on the distance to the set of separable states when the state becomes highly extendible, especially if its extendibility depends on the dimension.  In particular, we are interested how far can our family of k-hybrid be from the set of all separable states when $k$ is large. 

Setting $k = \lfloor f(n)\rfloor$, where $f(n)$ is the function of the sublinear growth, the state~\eqref{next} provides the the family of states that are $f(n)$-hybrid, with vanishing amount of entanglement on a single copy level in the limit of large $n$. The tensor product $\left(\eta^{AA'BB'}_{\lfloor f(n)\rfloor}\right)^{\otimes n}$ is an $f(n)$-hybrid state, which is very far from the set SEP of separable states.
\begin{lemma}\label{abovelemma}
Consider the function of sublinear growth $f(n)=o(n)$, and the state $\left(\eta^{AA'BB'}_{\lfloor f(n)\rfloor}\right)^{\otimes n}$. 

Then
\begin{enumerate}[(a)]
\item  $\displaystyle\min_{\sigma\in\mbox{SEP}}\norm{\eta^{AA'BB'}_{\lfloor f(n)\rfloor}-\sigma}_1\ge \frac{p_s}{\lfloor f(n)\rfloor}$, $p_s\in\left(\frac{1}{32},\frac{1}{28}\right)$\label{itemone}.
\item $\displaystyle\min_{\sigma^{(n)}\in\mbox{SEP}}\norm{\left(\eta^{AA'BB'}_{\lfloor f(n)\rfloor}\right)^{\otimes n}-\sigma^{(n)}}_1\to 1, \mbox{ as } n\to\infty.$\label{itemtwo}
\end{enumerate}
\end{lemma}
{\bf Proof}: 
The lower bound in~\eqref{itemone} follows from the observation that using the protocol described after Lemma~\ref{distillation}, Alice and Bob can distill the amount of entanglement, which equals to the lower bound for all values of $p_s$ in the range. The latter ensures that the state has PPT.

To prove~\eqref{itemtwo} we apply the result from~\cite{beigi_approximating_2010}, where authors derived the following lower bound for the trace distance from separable states to states, which are bound entangled and have PPT:
\be
\min_{\sigma^{(n)}\in\mbox{SEP}}\norm{\left(\eta^{ABA'B'}_{\lfloor f(n)\rfloor}\right)^{\otimes n}-\sigma^{(n)}}_1\ge 1-r_n,
\ee 
$r_n=(6\epsilon^{-1}+1)(\delta_n + 2d\frac{n}{n+n^2})+\epsilon^{\prime}$, where $d$ is the dimension of the underlying Hilbert space of $\eta^{ABA'B'}_{\lfloor f(n)\rfloor}$, which does not increase with $f(n)$. The terms $\delta_n, \epsilon^\prime$ indicate the speed of convergence of the values obtained by the process of tomography on $\sigma^{(n)}$, and $\left(\eta^{ABA'B'}_{\lfloor f(n)\rfloor}\right)^{\otimes n}$ accordingly. They arise as a result of the application the law of large numbers while performing the typical state tomography, and we can assume  them to be $\delta_n = \frac{A}{n}, \epsilon^\prime = \frac{B}{n}$ for some constants $A,B$. Lastly, $\norm{\eta^{AA'BB'}_{\lfloor f(n)\rfloor}-\sigma}_1\ge \frac{p_s}{\lfloor f(n)\rfloor}=\epsilon$ is taken from the $f(n)$-extendibility condition that is present in the statement of the Lemma~\ref{abovelemma}. Putting everything together, we get:
\be
\lim_{n\to\infty}r_n= \lim_{n\to\infty}\left[(6f(n)+1)\left(\frac{A}{n} + \frac{2dn}{n+n^2}\right)+\frac{B}{n}\right]=0.
\ee{\qed}
The result of the Lemma can be seen from a different perspective -- by means of performing the task of entanglement distillation using an expanded class of operations. If instead of just allowing LOCC we allow non-entangling maps, we will be able to distill from $\eta^{ABA'B'}_{\lfloor f(n)\rfloor}$ -- the amount of entanglement, which vanishes as $n\to\infty$. However, the total distillable entanglement of the state $\left(\eta^{ABA'B'}_{\lfloor f(n)\rfloor}\right)^{\otimes n}$ increases under non-entangling maps as $O(n^\epsilon)$ for some $\epsilon>0$, thereby increasing the distance from the set of all separable states. Each individual state in the tensor product has ever smaller amount of entanglement which decreases sublinearly as $n$ increases, yet, the entanglement of the tensor product increases (also sublinearly). This is not the case if we allow a super linearly vanishing amount of entanglement on the single copy level as $n$ increases. One may think that by letting the state be $\lfloor f(n)\rfloor$-cloneable, as $f(n)$ increases with $n$, yet having tensor product far from the set of separable states contradicts with the complete hierarchy of separability criteria of~\cite{doherty_complete_2004} in the sense that the constructed state is highly cloneable, yet far from the set of separable states. This is not the case, because for every finite $n$, we simultaneously fix $f(n)$, and for some $t>0$ we will fail to construct the symmetric extension of the state $\eta^{AA'BB'}_{\lfloor f(n)\rfloor}$ to $f(n)+t$ parties, thereby showing that it is entangled.

\subsection{Fully extendible states beyond CJ states of the erasure channel }

Previously we were working with the states, which were only extendible on one of the parties that possesses the state. Here we look on the bound entangled states with PPT, which are equally extendible on both parties. We will see that they are obtained by a slight modification of the previous constructions, symmetrizing them in a suitable way. To achieve this, we replace the bound entangled state $\phi^{AA'BB'}$ in
\be
\rho^{AA'BB'} = \frac{1}{k}\phi^{AA'BB'}+\frac{k-1}{k}\mathbb{1}^{AA'}\otimes\sigma^{BB'}
\ee 
by a state which belongs to the class ${\cal I}_{BE}$ of bound entangled states, which are invariant with respect to the exchange of the subsystems.

The following Lemma provides a way to take a rather generic bipartite bound entangled state and turn it into a highly extendible state on both parties, which turns out to be $K$-hybrid, where $K=\half k(k-1)$.
\begin{lemma}\label{biextlemma}
May $\rho^{AB}\in {\cal I}_{BE}$. Then the state
\be\label{biextendible}
\zeta^{\bf A} = \frac{1}{K}\sum_{i, j=1,i<j}^{k}\rho^{A_iA_j}\otimes\sigma^{{\bf A}\backslash A_iA_j}
\ee
is $K$-hybrid with the reduced state $\zeta^{A_iA_j} = \tr_{{\bf A}\backslash A_iA_j}(\zeta^{\bf A})$,
where ${\bf A} = A_1...A_k$,  $\sigma^{{\bf A}\backslash A_iA_j}$ is the separable state.
\end{lemma}
{\bf Proof:}
Fix $i_0,j_0$: $A_{i_0}$=$A$, $A_{j_0}$=$B$. From the permutation invariance of $\rho^{AB}$ we have $\tr_{A}\rho^{AB} = \tr_B\rho^{AB}$. This fact and the overall compositional symmetry of~\eqref{biextendible} shows that $\forall j: j\neq j_0$: $\zeta^{AB}=\zeta^{A_jB}=\tr_{{\bf A}\backslash \{A_{j},B\}}(\zeta^{\bf A})$. Similarly,  $\forall i: i\neq i_0$: $\zeta^{AB}=\zeta^{AA_i}=\tr_{{\bf A}\backslash \{A,A_i\}}(\zeta^{\bf A})$.
To show that $\zeta^{AB}$ is bound entangled, Bob performs measurement $M$ as in proof of Lemma~\ref{distillation} on his share of state $\sigma$, and communicates the result back to Alice. Now, parties share the state $\rho^{AB}$ with probability $\frac{1}{K}$,  which is bound entangled. $\qed$

The result of Lemma~\ref{biextlemma} can be directly generalized to the multipartite case. 

One could generate examples of permutationally invariant states from the general bipartite bound entangled states -- not necessarily permutationally invariant -- symmetrizing them with respect to the additional subsystems, as illustrated in the example below.
\begin{examp}
Let $\mu^{AB}$ be a bipartite bound entangled state. Then the state $\widetilde\mu^{AA'BB'} = \half[|10\ra\la10|^{A'B'}\otimes\mu^{AB}+ |01\ra\la01|^{A'B'}\otimes (F\mu^{AB}F)]$, where $F$ is a flip operator, belongs to the set ${\cal I}^{AB}$. Take $A_{i_0}$=$A$, $A_{j_0}$=$B$, and consider the state 
\begin{align}
&\zeta^{AB}=\\&\tr_{{\bf A}\backslash AA'BB'}\left(\frac{1}{K}\sum_{i,j=1,i<j}^{k}\widetilde\mu^{A_iA'_iA_jA'_j}\otimes\sigma^{{\bf A}\backslash A_iA_j}\right),
\end{align}
where $K=\half k(k-1)$. To see that this is bound entangled $K$-hybrid state, Alice and Bob perform the protocol as in the proof of Lemma~\ref{biextlemma}, followed by the projective measurement on a classical register $A'B'$. Depending on the outcome of their measurements, they will share either $\mu^{AB}$ or $F\mu^{AB}F$, which is bound entangled.
\end{examp}

\section{ Nontrivial zero-capacity channels with no known superactivation protocol}\label{apps}

The channels whose CJ states are k-hybrid provide an interesting case of zero-capacity channels, whose quantum capacity cannot be superactivated. The latter effect, originally discovered by~\cite{smith_superact} and later generalized by~\cite{brandao_when_2012}, consists of having two channels $\chn_1$, $\chn_2$ which are too noisy to transmit quantum information when used individually but when used together have positive capacity. The first channel in the setup produces bound entangled states with PPT, and another one produces certain 2-extendible states. More formally, superactivation is concisely described by the following relation:
\be
\begin{cases}
{\cal Q}(\chn_1)=0\\
{\cal Q}(\chn_2)=0\\
{\cal Q}(\chn_1\otimes\chn_2)>0.
\end{cases}
\ee

Consider a channel $\chn_H$ whose CJ state is k-hybrid  of the form~\eqref{pditkext}. The output of the channel is a PPT bound entangled state, which at the same time is $k$-extendible. Such channel retains the characteristic properties of the pdit channel $\chn_\pdit$~\cite{brandao_when_2012}. Also, this is the first known channel whose CJ state belongs to the set of PPT states, which are simultaneously k-extendible states. One can view this channel as the erasure channel which erases the input state with probability $1-\frac{1}{k}$. Previously, one could only construct a channel, which is binding entangled and whose CJ state belongs to the set of PPT 2-extendible states~\cite{doherty_complete_2004}. 

It turns out that $\chn_H$ cannot be activated using any of the channels that were previously instrumental in all of the protocols for superactivation up to date.

\begin{lemma}
Consider $\chn_H$ that produces states of the form~\eqref{pditkext}. 

Then
\be
\begin{cases}
{\cal Q}(\chn_H)=0\\
{\cal Q}(\chn_H\otimes \chn_\pdit)=0\\
{\cal Q}(\chn_H\otimes\chn_e^p)=0, p\in[\half,1)
\end{cases}
\ee
\end{lemma}
{\bf Proof:}
The first equality follows directly from its membership in the intersection of the set of two zero-capacity channels.

Taking $\chn_1 = \chn_H\otimes \chn_\pdit$ is again a channel that produces PPT states, hence, ${\cal Q}(\chn_1)=0$. Similarly, $\chn_2 = \chn_H\otimes\chn_e^p$ could be represented as another erasure channel with probability of erasure larger than $p$. Therefore, ${\cal Q}(\chn_2)=0$. \squareforqed

This curious case of hybrid channels leaves the question of superactivation of such class of channels open. This means that if they can be activated in principle, the activating channel will belong to a completely new class of zero-capacity channels, dissimilar to all zero-capacity channels we know now.

\section{Low-dimensional analogue of hybrid states}\label{classanalogue} 
All the constructions of the hybrid states pose a challenge to implement in the lab, because they require high dimensional Hilbert space to exist. It is also known that in $2\otimes2$ dimensions hybrid states as we known them cannot exist~\cite{horodecki_separability_1996}. We find a low-dimensional analogue to the hybrid states, which turn out to be 2-extendible and key-distillable with the two-way communication. The existence of such an analogue can be easily seen in the dimensions where there exist hybrid states, as each 2-hybrid state gives rise to extendible state with classical key. It is easy to see that such states exist in higher dimensions, as follows from the constructions of the k-hybrid states. Using the state~\eqref{erlike} , one can construct the $4\otimes5$ state, which contains a key and is k-extendible 
\be
\rho_{min}^{AA'BB'} = \frac{1}{k}\gamma_{min}^{AA'BB'} + \frac{k-1}{k}\mathbb{1}^{AA'}\otimes\sigma^{BB'},
\ee
where $\gamma_{min}^{AA'BB'}$ is the $4\otimes4$ pbit with the smallest possible dimensions of the shield recently introduced by~\cite{pankowski_low-dimensional_2011}, and $\sigma^{BB'}$ is the erasure flag.
The authors in~\cite{doherty_complete_2004} exhibit a $3\otimes3$ state which achieves the same goal, although there is no known distillation protocol to obtain a key. 

From the above considerations it is not possible to construct $2\otimes2$ states which achieve the goal, because for the systems on ${\cal H}_m\otimes {\cal H}_n$, where $mn\le6$, having PPT implies separability of the state~\cite{horodecki_separability_1996}. Therefore, if we lift the requirement for the bipartite state to have PPT, it turns out that it is possible to have a bipartite $2\otimes 2$ state that is 2-extendible and has two-way key (denoted as ($K_{\leftrightarrow}$). The latter means that if parties possess many copies of the state, they are able to distill the key using bidirectional classical communication channel. The existence of such states is not at all obvious, and even to certain extent paradoxical, as it passes the second test from the separability hierarchy, and yet contains two-way key.

Here we show existence of states $\rho^{AB}\in{\cal B}({\cal H}_2\otimes{\cal H}_2)$ that are extendible and for which $K_{\leftrightarrow}(\rho^{AB})>0$. We turn to the class of Bell-diagonal states, for which necessary and sufficient conditions for extendibility have recently been discovered~\cite{myhr_symmetric_2009}. More formally, let
\be\label{lowstate}
\rho^{AB} = \lambda_1\Phi^+ +  \lambda_2\Phi^-+ \lambda_3\Psi^++ \lambda_4\Psi^-,
\ee 
where $\sum_{i=1}^4\lambda_i=1$, and $\Phi^\pm, \Psi^\pm$ are Bell states. Put $\a_1 = \lambda_1-\lambda_2-\lambda_3+\lambda_4, \a_2=\sqrt{2}(\lambda_1-\lambda_4), \a_3 = \sqrt{2}(\lambda_2-\lambda_3)$. Then, $\rho^{AB}$ is extendible if and only if any of the following inequalities are satisfied~\cite{myhr_symmetric_2009}:
\begin{align}\label{extendibilitycritst}
&4\a_1(\a_2^2-\a_3^2)-(\a_2^2-\a_3^2)^2-4\a_1^2(\a_2^2+\a_3^2)\ge 0,\\
&\a_2^2-\a_3^2-2\sqrt{2}\a_1|\a_2|\ge 0,\\
&\a_3^2-\a_2^2+2\sqrt{2}\a_1|\a_3|\ge 0.\label{extendibilitycritend}
\end{align}
For the states expressed in Bell basis there also exists a sufficient condition for the distillation of key using single-copy measurements plus classical two-way processing protocols, expressed in terms of $\{\lambda_i\}_{i=1}^4$ ~\cite{acin_secrecy_2006}. It states that by sharing many copies of the state $\rho^{AB}$, one can distill secret key if and only if
\be\label{sufficientkey}
(\lambda_{max}-\lambda_{min})^2>(1-\lambda_{max}-\lambda_{min})(\lambda_{max}+\lambda_{min}),
\ee
where $\displaystyle\lambda_{max} = \max_i{\lambda_i}, \lambda_{min}=\min_i{\lambda_i}$.
\begin{obs}
Systems of inequalities~\eqref{extendibilitycritst}-\eqref{extendibilitycritend} and~\eqref{sufficientkey} are simultaneously satisfiable. The state~\eqref{lowstate} with eigenvalues satisfying these systems of inequalities are 2-extendible with $K_{\leftrightarrow}(\rho^{AB})>0$.
\end{obs}
The plot below describes the triples of eigenvalues $(\lambda_1, \lambda_2, \lambda_3)$ which are compatible with both systems of inequalities.
\begin{figure}[H]
\centering
\includegraphics[scale=0.64]{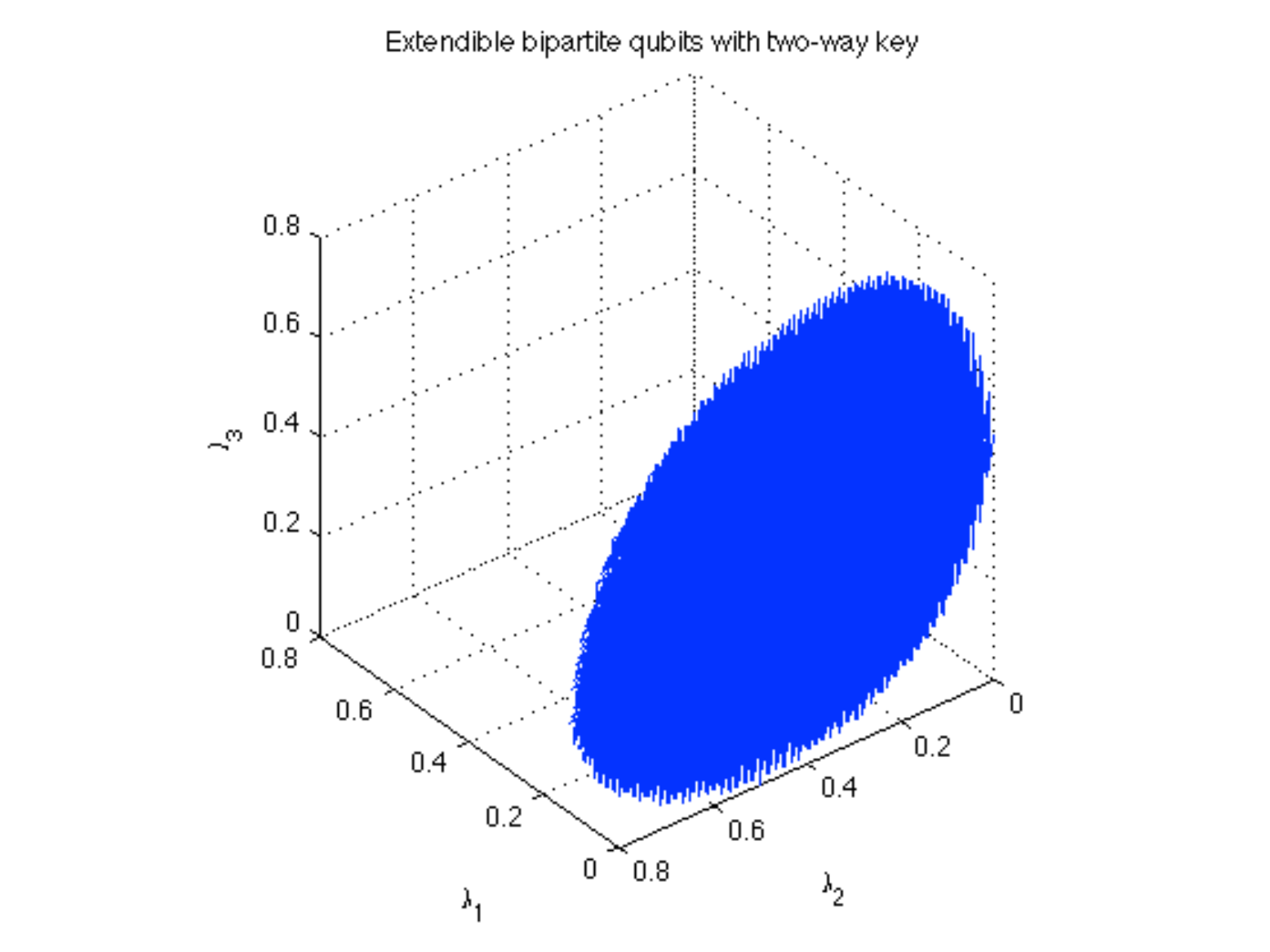}
\label{kext2waykey}
\end{figure}
\section{Conclusions}
We show the first explicit construction of the entangled states with PPT which are $k$-extendible for any fixed $k$, and explored their properties. In particular, we showed the existence of highly extendible states which are far from the set of separable states. We provided a simple way to distill entanglement when the parties share a hybrid state, which is based on pbits, by giving them a set of non-entangling resources in the form of backwards classical channel and backwards erasure channel, which has zero-capacity. 

The method used to construct hybrid states was inspired by the superactivation phenomenon. The corresponding hybrid channel has zero quantum capacity, and  it enables the parties to share an hybrid state extendible to any parties. An important open problem is to find another zero-capacity channel, which when used jointly with the hybrid channel will superactivate its capacity. It is evident that if such a channel exists, it must neither come from the set of channels, which produce bound entangled states with PPT, nor the set of channels whose CJ states are symmetrically extendible. Any such channel must be qualitatively different from all of the zero-capacity channels known to date, and thus help us to better understand the phenomenon of superactivation. Moreover, this problem is intimately related to the hard long-standing open question of the existence of the bound entanglement with negative partial transpose (NPT), which resists the numerous attempts to solve it. Can it be that the CJ states of the new class of zero-capacity channels, which are potent to superactivate the capacity of the hybrid channel are the examples of the elusive bound entangled states with NPT?\\\\
{\bf Acknowledgements:}
S.S. would like to thank Trinity College, Cambridge for their support throughout his PhD studies. J.O. acknowledges the support of the Royal Society.

\bibliographystyle{abbrv}

\end{document}